\documentclass[letterpaper]{article}

\usepackage[T1]{fontenc}

\usepackage{geometry}
\usepackage{setspace}
\usepackage{soul}
\setstcolor{red}

\usepackage[style = chem-acs]{biblatex}
\usepackage{graphicx}
\usepackage{float}
\newfloat{scheme}{htbp}{los}
\floatname{scheme}{Scheme}
\floatname{chart}{Chart}
\newfloat{graph}{htbp}{loh}

\usepackage{chemformula} 
\usepackage[version = 4]{mhchem} 

\usepackage{authblk}

\author[1,2]{Seungjun Lee}
\author[3]{Seung Gyo Jeong}
\author[2,3,4]{Jian-Ping Wang}
\author[3]{Bharat Jalan}
\author[2,4]{Tony Low}

\affil[1]{Department of Applied Physics, Kyung Hee University, Yongin 17104, Republic of Korea}
\affil[2]{Department of Electrical and Computer Engineering, University of Minnesota, Minneapolis, MN 55455, USA}
\affil[3]{Department of Chemical Engineering and Materials Science, University of Minnesota, Minneapolis, MN 55455, USA}
\affil[4]{School of Physics and Astronomy, University of Minnesota, Minneapolis, MN 55455, USA}

\title{Strain-Driven Altermagnetic Spin Splitting Effect in RuO$_2$}
\date{*Email: sjunlee@khu.ac.kr, tlow@umn.edu}

\begin{document}

\maketitle

\begin{abstract}
The non-relativistic spin-momentum locking in altermagnets gives rise to a time-reversal-odd spin Hall effect, known as the altermagnetic spin-splitting effect (ASSE). Although ASSE was first reported in RuO$_2$, subsequent experiments have yielded inconsistent results, leaving its spin-transport mechanism unclear. Here, we systematically investigate how strain, crystal orientation, and the Hubbard $U$ parameter influence the magnetic ground state and spin Hall response of RuO$_2$. Guided by recent experimental observations, we find that $U$ is likely smaller than the value required to induce intrinsic magnetism, suggesting that bulk RuO$_2$ and (001)/(101) RuO$_2$ thin films grown on TiO$_2$ are nonmagnetic in the absence of extrinsic effects. In contrast, (100) and (110) films exhibit strain-induced altermagnetic spin splitting, leading to a strong ASSE even without Hubbard $U$ corrections. These results reconcile previous experimental discrepancies and provide design guidelines for RuO$_2$-based spintronic devices.
\end{abstract}

\section*{Keywords}

RuO$_2$, Altermagnetism, Strain, Density-functional theory, Hubbard U correction, Spin Hall effect


Altermagnetism is a newly identified magnetic phase in which opposite spin densities are compensated by proper/improper rotations, mirrors, or nonsymmorphic operations (screw or glide), rather than by a pure lattice translation.~\cite{vsmejkal2020crystal,vsmejkal2022emerging,vsmejkal2022beyond}
This symmetry mechanism enables non-relativistic spin-momentum locking without net magnetization, creating new opportunities for spintronic and optoelectronic applications.~\cite{vsmejkal2020crystal,shao2021spin,zhou2021crystal,feng2022anomalous,gonzalez2021efficient,bose2022tilted,bai2022observation,weber2024all}
Among proposed material candidates, RuO$_2$ has attracted particular attention following two early findings: (i) neutron and x-ray scattering experiments reporting room-temperature magnetic order,~\cite{berlijn2017itinerant,zhu2019anomalous} and (ii) density functional theory (DFT) calculations with a Hubbard $U$ correction predicting strong spin splitting.~\cite{vsmejkal2022emerging,vsmejkal2022beyond,ahn2019antiferromagnetism}
These predictions motivated subsequent experiments reporting signatures consistent with altermagnetism, including anomalous and planar Hall effects~\cite{feng2022anomalous,tschirner2023saturation,song2025spin} and spin-resolved angle-resolved photoemission spectroscopy (ARPES).~\cite{fedchenko2024observation}

Despite these encouraging observations, the magnetic ground state of RuO$_2$ remains highly controversial. Large Hubbard $U$ values employed in early theory may spuriously favor (or overstabilize) a magnetic solution,~\cite{smolyanyuk2024fragility} and recent $\mu$SR measurements~\cite{hiraishi2024nonmagnetic,kessler2024absence} and follow-up ARPES, optical, and x-ray studies~\cite{liu2024absence,wenzel2025fermi,kiefer2025crystal} have all reported nonmagnetic behavior.
One likely reason for these disparate reports is that RuO$_2$’s magnetic ground state is highly sensitive to subtle changes in lattice parameters, which can be modified through crystallographic orientation, epitaxial strain, and film thickness.~\cite{jeong2024altermagnetic,jeong2025metallicity}

Adding to the uncertainty, spin-transport experiments have yielded signals that admit multiple interpretations. Early DFT calculations predicted that the altermagnetic spin-splitting effect (ASSE) in RuO$_2$ could generate an exceptionally large effective spin Hall angle ($\sim$28\%),~\cite{gonzalez2021efficient} raising hopes that ASSE detection would offer a decisive signature of altermagnetism. 
However, while several studies indeed reported finite ASSE signals,~\cite{bose2022tilted,bai2022observation,bai2023efficient,guo2024direct,liao2024separation,li2025fully,Jung2025} others found no ASSE.~\cite{wang2024inverse,wang2025robust} 
(Table S1 in Supplementary Information (SI))

Interpreting these results is further complicated by experimental details, such as substrate-induced strain and film thickness-dependence.~\cite{ji2013strain,quarterman2018demonstration}
In theoretical side, the appropriate Hubbard $U$ value for RuO$_2$ remains unsetteled.

To reconcile the ongoing debates surrounding RuO$_2$, we present a unified first-principles framework that treats epitaxial strain, crystallographic orientation, and electronic correlations on equal footing, and we track their coupled impact on both the magnetic ground state and spin-transport responses. We derive symmetry-constrained tensor forms for the spin Hall effect (SHE) and ASSE and analyze their angular dependence. We then investigate the electronic structure and spin-transport properties of (001), (100), (101), and (110) oriented films.
Our results show that even without the Hubbard $U$ correction, (100) and (110) oriented RuO$_2$ films can host magnetism and sizable ASSE. Finally, we discuss possible origins of the conflicting experimental observations.


\begin{figure}[h]
\includegraphics[width=0.50\columnwidth]{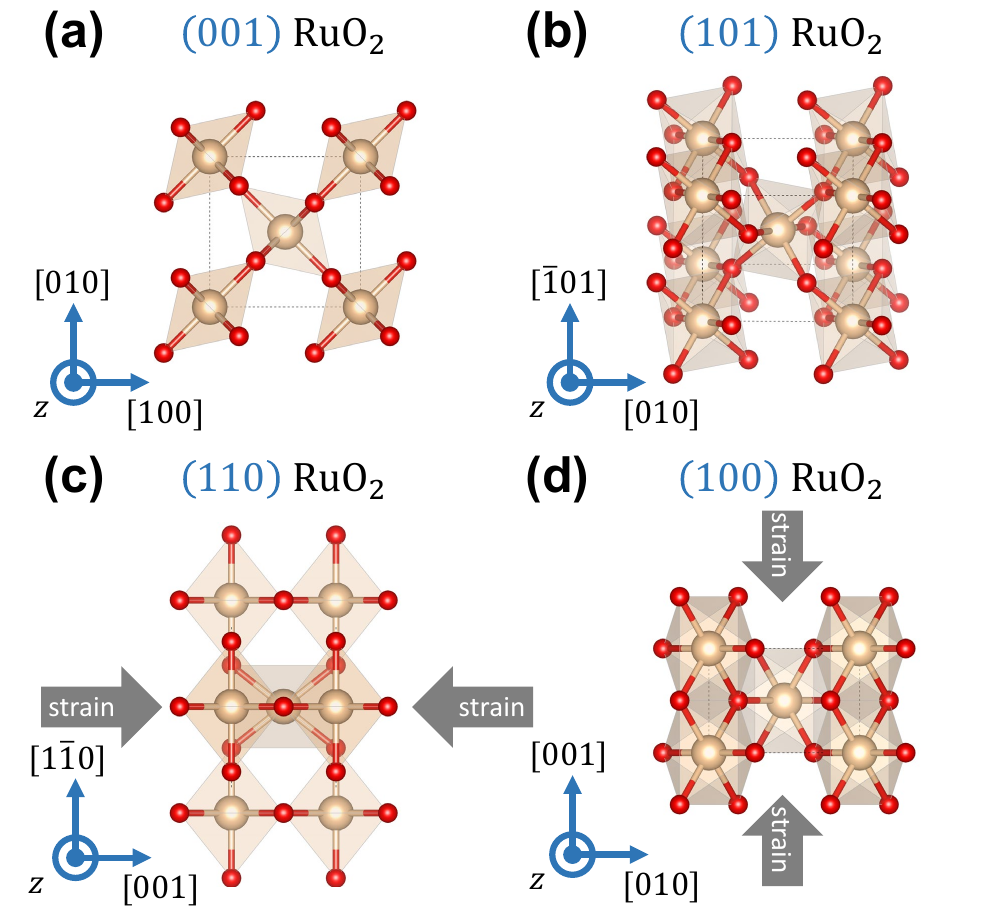}
\caption{(a-d) Top views of the crystal structures of RuO$_2$ with orientations (001), (101), (110), and (100), respectively. The arrows in (c) and (d) indicate the compressive strain imposed by the TiO$_2$ substrate along its [001] direction.
}\label{fig1}
\end{figure}

We begin with our discussion on the general symmetry requirements of both SHE and ASSE. 
Throughout this work, we denote them as ${\sigma}^{\gamma}_{\alpha \beta}$ and ${\sigma}^{\rm{A}, \gamma}_{\alpha \beta}$, respectively, where $\alpha$, $\beta$, and $\gamma$ represent cartesian coordinates and the superscript A indicates ASSE.
Within the constant-broadening (constant-$\Gamma$) model, ${\sigma}^{\gamma}_{\alpha \beta}$ and ${\sigma}^{\rm{A}, \gamma}_{\alpha \beta}$ can be interpreted as Fermi sea and Fermi surface contributions of a linear response of spin-current operator $j^{\gamma}_{\alpha}=\frac{1}{2}\{v_{\alpha} , s_{\gamma}\}$ described as,~\cite{kubo4,kubo1}
\begin{equation}
\sigma_{\alpha \beta}^{\rm{A}, \gamma} = \displaystyle -\frac{e \hbar}{\pi}  \sum_{\textbf{k}nm} \frac{{\Gamma}^2\operatorname{Re}
[\langle n\textbf{k}|j_{\alpha}^{\gamma}|m\textbf{k} \rangle \langle m\textbf{k}| v_{\beta}| n \textbf{k}\rangle]}{[({\epsilon}_{F} - {\epsilon}_{n{\textbf{k}}})^2+{\Gamma}^2][({\epsilon}_{F} - {\epsilon}_{m{\textbf{k}}})^2+{\Gamma}^2]},
\end{equation}
\begin{equation}
\sigma_{\alpha \beta}^{{\gamma}} = \displaystyle -2e\hbar \sum_{\textbf{k}n,m\neq n}(f_{n\textbf{k}} - f_{m\textbf{k}}) \frac{\operatorname{Im}[\langle n\textbf{k}|j_{\alpha}^{\gamma}|m\textbf{k} \rangle \langle m\textbf{k}| v_{\beta}| n\textbf{k}\rangle]}{({\epsilon}_{n\textbf{k}} - {\epsilon}_{m\textbf{k}})^2  + {\Gamma}^2},
\end{equation}
where $\Gamma$ indicate a broadening constant, $v$ and $s$ are velocity and spin operators, $f$ is a Fermi-Dirac distribution, and $| n\textbf{k}\rangle$ is the eigenstate associated with the band $\epsilon_{n\textbf{k}}$ of the unperturbed system.
It is importantly note that ${\sigma}^{\gamma}_{\alpha \beta}$ (${\sigma}^{\rm{A}, \gamma}_{\alpha \beta}$) is even (odd) under time reversal operator ($\mathcal{T}$).~\cite{kubo1,vzelezny2017spin} 
Therefore, ASSE also often referred to as the $\mathcal{T}$-odd SHC.
Intuitively, thus, ${\sigma}^{\rm{A}, \gamma}_{\alpha \beta}$ must be zero in nonmagnetic system, but can be non-zero in altermagnet. The shape of both $\mathcal{T}$-even and $\mathcal{T}$-odd SHC tensors are intrinsically constrained by materials' symmetries.~\cite{vzelezny2017spin,roy2022unconventional,jeong2025magnetic}

Figure~\ref{fig1}(a) show top views of (001) RuO$_2$ crystal structure and Table~\ref{table2} summarizes its symmetry-restricted form of the spin Hall conductivity (SHC) tensors assuming non-zero N\'eel vector ($\vec{N}$) aligned along [001].~\cite{bose2022tilted,vzelezny2017spin} 
Even without spin–orbit coupling (SOC), the altermagnetic spin splitting induces finite spin currents, such as ${\sigma}^{\rm{A}, z}_{xy}$. 
With SOC, lattice–spin coupling introduces conventional $\mathcal{T}$-even terms as well as extra $\mathcal{T}$-odd contributions, e.g., ${\sigma}^{\rm{A}, x}_{yz}$ and ${\sigma}^{\rm{A}, y}_{zx}$.
Given that the Fermi surface is predominantly polarized along [001], it is natural that ${\sigma}^{\rm{A}, z}_{xy}$ dominates over ${\sigma}^{\rm{A}, x}_{yz}$ and ${\sigma}^{\rm{A}, y}_{zx}$.
Consequently, in FM/(001) RuO$_2$ thin-film heterostructures, the dominant ASSE contribution is ${\sigma}^{\rm{A}, z}_{xy}$. 
However, this component is difficult to detect with conventional spin-torque ferromagnetic resonance or second-harmonic measurements, because these techniques primarily probe spin currents flowing along out-of-plane directions. For this reason, alternative crystal orientations are often employed in experiments.

\begin{table}[b]
\centering
\caption {Symmetry-restricted shape of spin Hall conductivity tensors (SHC) of RuO$_2$ with [001] N\'eel vector. ${\sigma}^{\gamma}_{\alpha \beta}$ and ${\sigma}^{\rm{A}, \gamma}_{\alpha \beta}$ represent $\mathcal{T}$-even and $\mathcal{T}$-odd SHC components, respectively. Since the response considered here is the DC limit, all tensor elements are real-valued.
\label{table2}}
\begin{tabular}{c|c|c} 
\hline
\hline
 & no SOC & SOC \\
\hline
$\sigma^x$ & 
$\begin{pmatrix} 0 & 0 & 0 \\ 0 & 0 & 0  \\ 0 & 0 & 0 \end{pmatrix}$  
& 
$\begin{pmatrix} 0 & 0 & 0 \\ 0 & 0 & {\sigma}^{x}_{yz}+{\sigma}^{\rm{A}, x}_{yz}  \\ 0 & -{\sigma}^{y}_{zx}+{\sigma}^{\rm{A}, y}_{zx} & 0 \end{pmatrix}$ \\
$\sigma^y$ & $\begin{pmatrix} 0 & 0 & 0 \\ 0 & 0 & 0  \\ 0 & 0 & 0 \end{pmatrix}$  
& $\begin{pmatrix} 0 & 0 & -{\sigma}^{x}_{yz}+{\sigma}^{\rm{A}, x}_{yz} \\ 0 & 0 & 0  \\ {\sigma}^{y}_{zx}+{\sigma}^{\rm{A}, y}_{zx} & 0 & 0 \end{pmatrix}$ \\
$\sigma^z$ & $\begin{pmatrix} 0 & {\sigma}^{\rm{A}, z}_{xy} & 0 \\ {\sigma}^{\rm{A}, z}_{xy} & 0 & 0  \\ 0 & 0 & 0 \end{pmatrix}$
& $\begin{pmatrix} 0 & {\sigma}^{z}_{xy}+{\sigma}^{\rm{A}, z}_{xy} & 0 \\ -{\sigma}^{z}_{xy}+{\sigma}^{\rm{A}, z}_{xy} & 0 & 0  \\ 0 & 0 & 0 \end{pmatrix}$ \\
\hline
\hline
\end{tabular}
\end{table}

Figure~\ref{fig1}(b-d) visualize crystal structures of (101), (110), and (100) RuO$_2$ crystal structures.
Neglecting additional symmetry breaking from epitaxial strain, the SHC tensors of a rotated RuO$_2$ crystal $(\sigma')$ can be obtained from its original tensors $(\sigma)$ using the rotation matrix $D$ as follows ${\sigma'}^{k}_{ij}=\sum_{lmn}D_{il}D_{jm}D_{kn}{\sigma}^{n}_{lm}$ where $ijk$ and $lmn$ are cartesian axes of the rotated and original RuO$_2$.~\cite{bose2022tilted} 
For (101) RuO$_2$, $\vec{N}$ is tilted from out-of-plane direction by arctan($c/a$)$\sim$~35$^{\circ}$, where $a$ and $c$ are lattice constants of RuO$_2$ along [100] and [001] directions. Therefore, conventional and unconventional $\mathcal{T}$-even (${\sigma}^{y}_{zx}$, ${\sigma}^{z}_{zx}$) and $\mathcal{T}$-odd SHC terms (${\sigma}^{\rm{A}, y}_{zx}$, ${\sigma}^{\rm{A}, z}_{zx}$) arise. This is particularly interest because non-zero unconventional SHC facilitate field-free switching in spin-orbit torque devices.~\cite{macneill2017control,liu2021symmetry,yang2025large}
In contrast, for both (110) and (100) RuO$_2$, $\vec{N}$ lies within the plane, and the dominant ASSE components are ${\sigma}^{\rm{A}, y}_{xx}$ and ${\sigma}^{\rm{A}, y}_{zx}$, respectively. 
Moreover, the lattice mismatch between RuO$_2$ and TiO$_2$ induces a strong compressive and tensile strain in RuO$_2$ thin films along [001] ($-4.7\%$) and [100]=[010] ($+2.2\%$) directions, respectively.~\cite{jeong2024altermagnetic,jeong2025metallicity}
These strain effects are significant in (001), (100), and (110) RuO$_2$, but are relatively weak in (101) RuO$_2$ ($+0.2\%$ for [10$\bar{1}$]).
Such strain can significantly modify the electronic structure and SHC tensors of RuO$_2$ and must be carefully considered when analyzing experimental results. 

\begin{figure}[h]
\includegraphics[width=0.50\columnwidth]{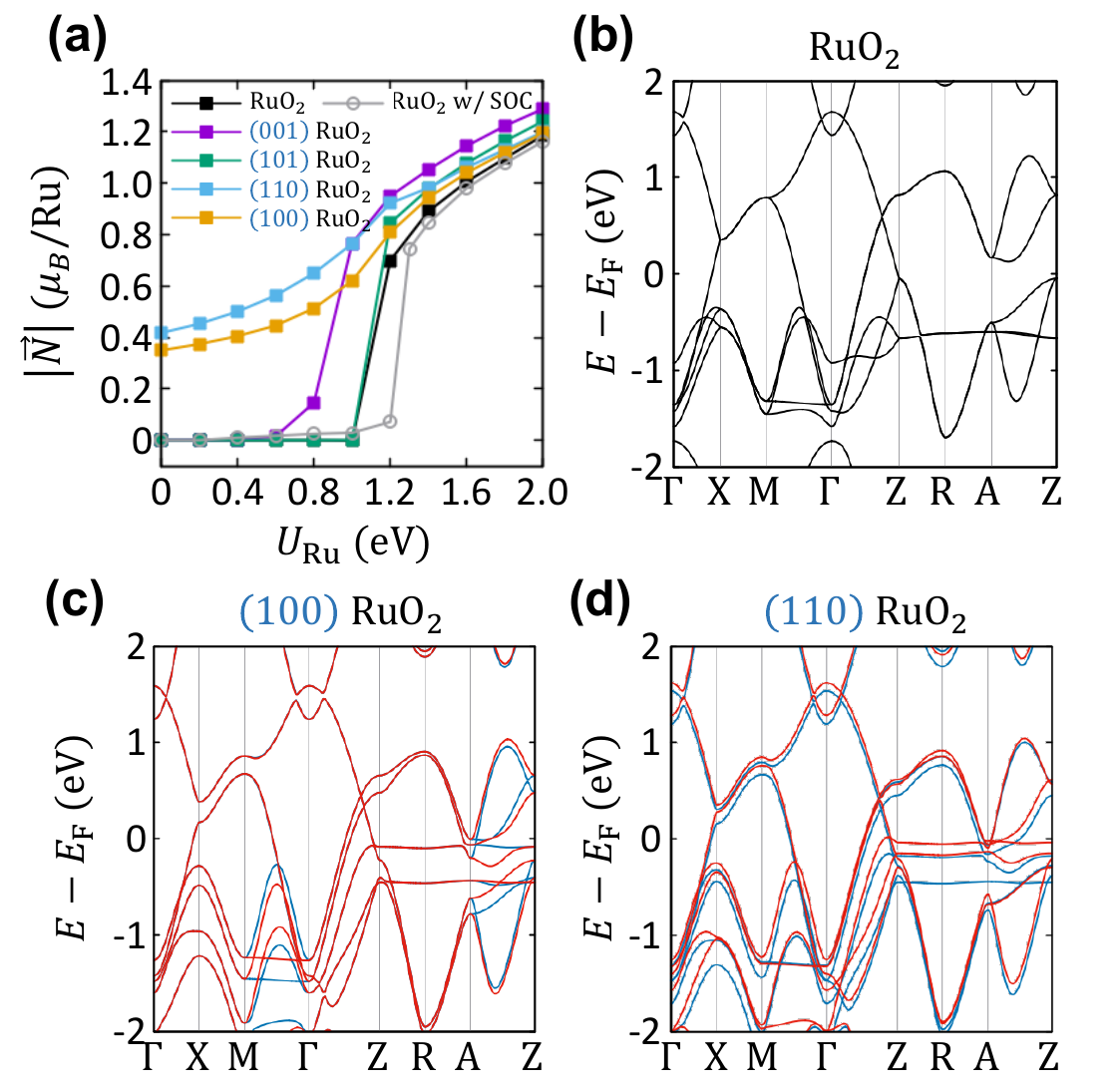}
\caption{(a) Hubbard $U$ dependent N\'eel vector magnitude of bulk RuO$_2$, and fully-strained (001), (101), (110) and (100) RuO$_2$ on TiO$_2$ substrates, respectively, without spin-orbit coupling (SOC), together with the result for bulk RuO$_2$ with SOC. (b-d) Electronic structures of bulk RuO$_2$, fully-strained (100) and (110) RuO$_2$, without $U$ and SOC.
}\label{fig2}
\end{figure}

To investigate the $U$- and strain-dependent electronic structure of RuO$_2$, we performed first-principles DFT calculations. To model epitaxial strain effects from the TiO$_2$ substrate, the in-plane lattice constants of RuO$_2$ were fixed to those of bulk TiO$_2$, while the atomic positions and out-of-plane lattice constant were fully relaxed within DFT calculation. Hereafter, we refer to fully strained RuO$_2$ on ($hkl$) TiO$_2$ as ($hkl$) RuO$_2$ for convenience. To quantify the local magnetic moment, we defined the N\'eel vector magnitude as
$|\vec{N}| = |\vec{\mu}_{\text{Ru}_1} - \vec{\mu}_{\text{Ru}_2}|/2$,
where $\vec{\mu}_{\text{Ru}_1}$ and $\vec{\mu}_{\text{Ru}_2}$ denote the local magnetic moments of the two Ru atoms in the formula unit. All other computational details are provided in SI.

Figure~\ref{fig2}(a) summarizes the Hubbard $U$-dependent local magnetic moments of bulk, (001), (101), (110) and (100) RuO$_2$, respectively. Without SOC, a critical Hubbard $U$ value for nonmagnetic to magnetic phase transition is predicted between 1.0 and 1.2~eV for Ru $4d$ in a good agreement with the previous DFT calculation.~\cite{smolyanyuk2024fragility} Including SOC slightly increases the critical $U$ value ($\sim$1.2~eV). The (101) RuO$_2$ case shows nearly the same critical $U$ as the bulk RuO$_2$, due to weak lattice mismatch.
For (001) RuO$_2$, the critical $U$ is slightly decreased ($\sim$0.8~eV) due to the reduced out-of-plane lattice constant induced by the Poisson ratio.~\cite{Forte2025} Interestingly, both (110) and (100) RuO$_2$ exhibit finite $|\vec{N}|$ even with zero Hubbard $U$ value. Notably, once $U>1.2$~eV, $|\vec{N}|$ becomes insensitive to strain. This implies that, in RuO$_2$, the origin of local magnetism also depends on the Hubbard $U$ parameter. In the absence of $U$, itinerant magnetism arises from Fermi surface instability associated with the high density of states which is strain-sensitive, while large $U$ drives strain-independent localized magnetism.~\cite{Forte2025}

Figure~\ref{fig2}(b-d) show electronic band structures of bulk, (100), and (110) RuO$_2$, respectively, without SOC and Hubbard $U$.
We clearly found the altermagnetic spin splitting in (110) and (100) RuO$_2$, while overall band structure highly resembles that of bulk RuO$_2$.
Spin splitting energies in (110) and (100) RuO$_2$ are order of a few tens of meV, which is much smaller than the predicted value ($\sim$1~eV) using $U=2$~eV.~\cite{vsmejkal2022emerging,jeong2025magnetic}
In (110) RuO$_2$, in-plane strain along [1$\bar{1}$0] direction further lowers its symmetry and thus lifts spin-degeneracy over the entire momentum space, indicating that it becomes non-compensated ferrimagnet instead of ideal altermagnet.~\cite{jeong2025metallicity,Forte2025}
Additional DFT calculations confirm that finite magnetism in RuO$_2$ emerges only beyond a critical epitaxial strain threshold, above which  (110) RuO$_2$ develops a finite $|\vec{m}|$ while (100) RuO$_2$ remains fully  compensated, consistent with their distinct symmetry properties (see Fig.~S1 and~S2 in SI).

\begin{figure*}[h]
\includegraphics[width=1.00\columnwidth]{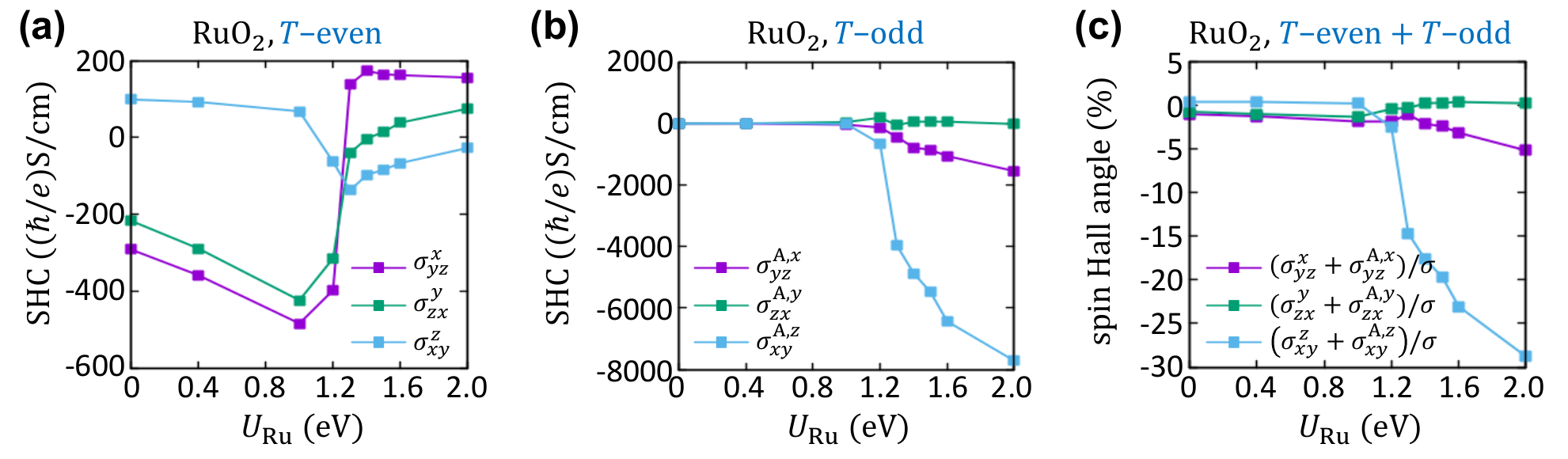}
\caption{(a-c) Hubbard $U$ dependent $\mathcal{T}$-even, $\mathcal{T}$-odd spin Hall conductivities, and spin Hall angles of strain-free bulk RuO$_2$ at Fermi level. $x||[100]$, $y||[010]$, and $z||[001]$ cartesian coordinates indicate corresponding directions.
}\label{fig3}
\end{figure*}

The above discussion emphasize that determining a reasonable size of Hubbard $U$ in RuO$_2$ is crucial. Recent discussions in the research community point toward the conclusion that a large $U>1.2$~eV is unlikely, as supported by multiple experimental observations.
First, the experimentally measured $|\vec{\mu}_{\rm{Ru}}|$ of 0.05$\sim$0.15~$\mu_{\rm B}$ is an order of magnitude smaller than the $\sim$1.0~$\mu_{\rm B}$ predicted by DFT calculations with $U>1.2$~eV.~\cite{berlijn2017itinerant,zhu2019anomalous}
Second, the observed absence of magnetism in neutron experiments on high-purity bulk RuO$_2$ further supports a small or zero $U$.~\cite{hiraishi2024nonmagnetic,kessler2024absence}
Finally, recent optical spectroscopy measurements~\cite{wenzel2025fermi,jeong2025anisotropic} and ARPES spectra~\cite{liu2024absence,Yichen2025} are more consistent with $U \sim 0$~eV rather than $U>1.2$~eV.
Furthermore, an emergence of mirror-even spin splitting in a 2~nm RuO$_2$ film on a (110) TiO$_2$ substrate~\cite{Yichen2025} is fully consistent with DFT results of strain-induced magnetism without Hubbard $U$ correction.
Nonetheless, it is important to note that a small but finite $U$ cannot be completely ruled out, thus surveying $U$-dependent properties remains valuable for a comprehensive understanding of RuO$_2$.

Figure~\ref{fig3}(a) shows the $U$-dependent $\mathcal{T}$-even SHCs of bulk RuO$_2$ at the Fermi level ($E_{\rm{F}}$). At $U=0$, ${\sigma}^{x}_{yz}$, ${\sigma}^{y}_{zx}$, and ${\sigma}^{z}_{xy}$ are calculated to be $-292$, $-217$, and $99$ ($\hbar$/e)S/cm, respectively, in a good agreement with the previous calculation using the same method.~\cite{PhysRevB.95.235104} By considering its experimental charge conductivity of $\sigma=2.8{\times}10^{4}~{\Omega}^{-1}\rm{cm}^{-1}$ at room temperature,~\cite{ryden1970electrical} the corresponding spin Hall angle (SHAs) is order of 0.35$\sim$1~{\%}.
The obtained $\mathcal{T}$-even SHCs are not very sensitive either $U<1.0$~eV or $U>1.4$~eV. However, near $U{\sim}1.2$~eV, they abruptly evolves due to the magnetic phase transition. Overall, the magnitudes remain comparable to the $U=0$ case, suggesting that RuO$_2$ is not a promising spin–orbit torque material when only $\mathcal{T}$-even contributions are considered.

Figures~\ref{fig3}(b) and (c) shows $U$-dependent $\mathcal{T}$-odd SHCs and SHAs, respectively.
As increasing $U$, altermagnetic spin splitting gives rise to $\mathcal{T}$-odd SHCs which explosively increase near $U{\sim}$1.2~eV.
With $U=2$~eV, the dominant $\mathcal{T}$-odd SHC term, ${\sigma}^{\rm{A}, z}_{xy}$, is calculated to be $-7717$~($\hbar$/e)S/cm, which nearly four times lager than SHC of Pt ($\sim$2000~($\hbar$/e)S/cm).~\cite{guo2008intrinsic} 
The corresponding Hall angle, ${\sigma}^{\rm{A}, z}_{xy}/{\sigma}=-28\%$, is in good agreement with a previous study.~\cite{gonzalez2021efficient} 
However, this value relies on the unrealistically large $U=2$~eV. 
We further note that both $\mathcal{T}$-even and $\mathcal{T}$-odd SHCs are sensitive to $E_{\rm{F}}$, and the detailed dependencies are summarized in Figs. S3 and S4 in SI.

\begin{figure*}[h]
\includegraphics[width=1.00\columnwidth]{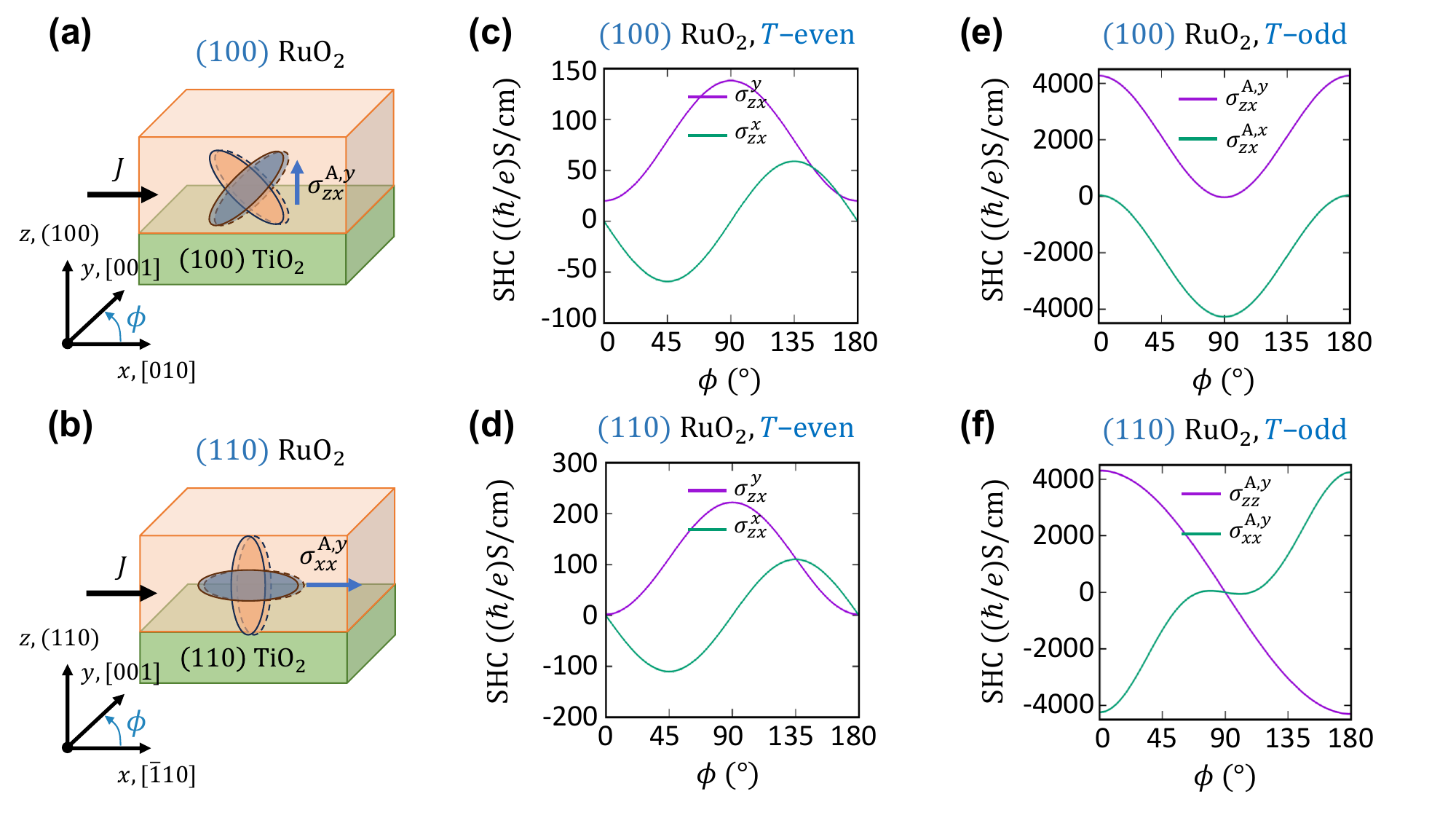}
\caption{
(a, b) Schematics of (100) and (110) RuO$_2$/TiO$_2$ heterostructures, respectively. 
The charge current $J$ direction is rotated in the film plane by an azimuthal angle $\phi$ with respect to the crystallographic x axis.
The Fermi surfaces illustrated in the schematics represent the orientation of the 
altermagnetic spin splitting for each device geometry, with blue and orange colors 
denoting opposite spin characters.
The representative (c, d) $\mathcal{T}$-even and (e, f) $\mathcal{T}$-odd SHC tensor components in (100) and (110) RuO$_2$, without Hubbard $U$ correction.
}\label{fig4}
\end{figure*}

Through the previous discussion, we can reasonably argue that including Hubbard $U$ may overestimate $\mathcal{T}$-odd SHC.
Since (100) and (110) RuO$_2$ exhibit magnetism even without $U$, we evaluate their representative $\mathcal{T}$-even and $\mathcal{T}$-odd SHC tensor components, summarized in Fig.~\ref{fig4}. Here, we represented their azimuthal angle ($\phi$, in-plane rotation) dependence to discuss how we differentiate these two contributions in experiments, as shown in Fig.~\ref{fig4}(a) and (b). The entire shape of $\mathcal{T}$-even and $\mathcal{T}$-odd SHC tensors of strained RuO$_2$ thin film is summarized in Tables S2 and S3 in SI.

As shown in Fig.~\ref{fig4}(c) and (d), the maximum order of $\mathcal{T}$-even SHCs in both (100) and (110) RuO$_2$ are a few hundred of ($\hbar$/e)S/cm, and ${\sigma}^y_{zx}$ reach maximum at ${\phi}=90^{\circ}$ where charge current along [001] direction. This angular dependence originates from anisotropy between [001] and [010] or [$\bar{1}10$] directions. We emphasize this clear angular dependence in $\mathcal{T}$-even SHE must be carefully considered in analyzing experimental data.
For both orientations, neither ${\phi}=0^{\circ}$ or ${\phi}=90^{\circ}$, unconventional component, ${\sigma}^x_{zx}$ becomes non-zero due to the lack of in-plane mirror operation corresponding to the charge current direction.~\cite{roy2022unconventional}

Figure~\ref{fig4}(e) shows $\phi$-dependent dominant $\mathcal{T}$-odd SHC of (100) RuO$_2$. At ${\phi}=0^{\circ}$, ${\sigma}^{\rm{A},y}_{zx}$ is calculated to be 4271~($\hbar$/e)S/cm and the corresponding SHA is around 15.3~\%, even without $U$.
This significant SHA make it a promising spin-orbit torque material, and it is comparable or even larger than those of notable heavy metals such as Pt (8\%)~\cite{PhysRevLett.106.036601} and Ta (15\%)~\cite{liu2012spin}.
With non-zero $\phi$, an unconventional component, 
At ${\phi}=90^{\circ}$, ${\sigma}^{\rm{A},x}_{zx}$ becomes dominant but ${\sigma}^{\rm{A},y}_{zx}$ goes to zero because the spin polarization direction is locked to N\'eel vector. 
Since this unconventional spin component is not generated by $\mathcal{T}$-even SHC at ${\phi}=90^{\circ}$, this term can play an important experimental evidence of altermagnetic order of (100) RuO$_2$.
Due to the rotated Fermi surface, the conventional $\mathcal{T}$-odd SHC term, ${\sigma}^{\rm{A},y}_{zx}$, is zero in (110) RuO$_2$. Instead, it can generate longitudinal spin current as similar with ferromagnet, as shown in Fig.~\ref{fig4}(f). At ${\phi}=0^{\circ}$, ${\sigma}^{\rm{A},y}_{xx}$ and ${\sigma}^{\rm{A},y}_{zz}$ are calculated to be $-4242$ and 4301~($\hbar$/e)S/cm, respectively. 
This longitudinal $\mathcal{T}$-odd SHC may be utilized in altermagnetic tunneling magnetoresistance applications.~\cite{shao2021spin,noh2025tunneling}

We note that, without interfacial effects, the Dzyaloshinskii-Moriya (DM) interaction vanishes in strained RuO$_2$ since the epitaxially strained space groups $Cmmm$ (No.~65) and $Pnnm$  (No.~58) for (110) and (100) RuO$_2$, respectively, both retain inversion symmetry. Our DFT calculations including SOC confirm that the magnetic easy axis remains along the [001] direction in both cases (See Fig.~S5 in SI). In experimental thin films, however, the RuO$_2$/TiO$_2$ interface breaks inversion symmetry, which could induce a finite interfacial DM interaction and tilt the N\'{e}el vector away from [001].~\cite{jeong2024altermagnetic} This can activate additional non-zero SHC tensor elements, offering an additional route to engineer spin transport properties in RuO$_2$-based heterostructures.

Finally, at this stage, we discuss the origin of the recent debate surrounding charge-to-spin conversion (or spin-to-charge conversion) experiments in ferromagnet/RuO$_2$ thin films, as summarized in Table S1 in SI. Our theoretical analysis shows that epitaxial strain alone is insufficient to stabilize magnetism in (101) RuO$_2$. Therefore, the previous experimental observations of ASSE in this orientation are likely influenced by extrinsic factors such as defects, doping,~\cite{smolyanyuk2024fragility} or surface/interface-induced magnetic moments.~\cite{Ho2025,akashdeep2025surface} 
Thus, any $\mathcal{T}$-odd SHC in (101) RuO$_2$ could be finite, but is expected to be small.
This explains why the observed $\mathcal{T}$-odd SHC in (101) RuO$_2$ is much smaller than the theoretical predictions based on large $U$ value.

Instead, we shows that (100) RuO$_2$ is a more promising platform for unambiguous detection of $\mathcal{T}$-odd SHC, because its $\mathcal{T}$-even and $\mathcal{T}$-odd components exhibit clearly distinct angular dependencies. This makes it particularly suitable for spin-torque and second-harmonic measurements. However, its experimental realization remains challenging due to the strong thickness dependence of epitaxial strain. Previous studies have shown that strain along the [001] direction is rapidly relaxed as increasing film thickness above 4~nm.~\cite{jeong2025metallicity,Forte2025} Therefore, to observe nonzero $\mathcal{T}$-odd SHC in (100) RuO$_2$, films must be sufficiently thin and exhibit high crystalline quality.
In addition, the sign of the $\mathcal{T}$-odd SHC should flip when the Néel vector is reversed. Therefore, the relative sign between the $\mathcal{T}$-odd and $\mathcal{T}$-even SHCs can be either positive or negative, depending on the experimental conditions. This sign change can serve as an experimental strategy to identify the presence of a $\mathcal{T}$-odd SHC.~\cite{Jung2025}

We further note that experimentally observed $\mathcal{T}$-odd SHC signals in (100) RuO$_2$ thin films are also significantly smaller than theoretical predictions,~\cite{bai2022observation,Jung2025} which we attribute to two main factors: strain relaxation in thicker films that suppresses the strain-induced magnetic order, and the multi-domain nature of experimental samples where contributions from domains with varying N\'{e}el vector orientations partially cancel each other. Therefore, achieving sufficiently strong epitaxial strain together with single-domain N\'{e}el vector alignment would be essential to experimentally access the intrinsic $\mathcal{T}$-odd SHC, making N\'{e}el vector switching and domain alignment important directions for future experimental research on RuO$_2$-based spintronic devices.


In summary, we have performed a comprehensive theoretical investigation of the SHE and ASSE in RuO$_2$, focusing on the influence of strain, crystallographic orientation, and electronic correlations. Our results demonstrate that the Hubbard $U$ parameter plays a critical role in determining the magnetic ground state and spin Hall responses. A large $U$ ($>1.2$ eV), often assumed in previous studies, unrealistically enhances magnetism and $\mathcal{T}$-odd SHC. In contrast, experimental evidence and our calculations consistently suggest that RuO$_2$ is weakly correlated, with small or negligible $U$.
Without $U$, we find that, while (001) and (101) oriented RuO$_2$ films remain nonmagnetic, (100) and (110) orientations can exhibit altermagnetic spin splitting.
In particular, strain-induced altermagnetism in (100) RuO$_2$ generates a remarkably large $\mathcal{T}$-odd SHC, corresponding to the SHA of approximately 15.3\%. More importantly, the $\mathcal{T}$-even and $\mathcal{T}$-odd contributions in this geometry exhibit distinct angular dependencies, providing an unambiguous experimental pathway to verify the ASSE. Our findings explain why ASSE may have been absent in some experiments on bulk-like or (101) films, while simultaneously predicting a strong, detectable signature in high-quality, ultra-thin (100) films where epitaxial strain is maximized.

\section*{Acknowledgements}

This work was supported by AFOSR sponsored MURI (Grant \#FA9550-25-1-0262) and NSF MEITY Program (ECCS 2415836). SL is also supported by a grant from Kyung Hee University (KHU-20262255).

\section*{Supporting information}

The following files are available free of charge.
\begin{itemize}
  \item Computational details; Strain-dependent magnetism of RuO$_2$; Altermagnetic symmetry of strained RuO$_2$; Hubbard $U$ and Fermi level dependent spin Hall conductivity tensor components; Magnetocrystalline anisotropy energy of strained RuO$_2$; Summary of reported charge-to-spin (and spin-to-charge) conversion experiments in RuO$_2$; $\mathcal{T}$-even and $\mathcal{T}$-odd spin Hall conductivity tensors
\end{itemize}

\printbibliography

@article{vsmejkal2020crystal,
  title={Crystal time-reversal symmetry breaking and spontaneous Hall effect in collinear antiferromagnets},
  author={{\v{S}}mejkal, Libor and Gonz{\'a}lez-Hern{\'a}ndez, Rafael and Jungwirth, Tom{\'a}{\v{s}} and Sinova, Jairo},
  journal={Sci. adv.},
  volume={6},
  number={23},
  pages={eaaz8809},
  year={2020},
  publisher={American Association for the Advancement of Science}
}

@article{vsmejkal2022emerging,
  title={Emerging research landscape of altermagnetism},
  author={{\v{S}}mejkal, Libor and Sinova, Jairo and Jungwirth, Tomas},
  journal={Phys. Rev. X},
  volume={12},
  number={4},
  pages={040501},
  year={2022},
  publisher={APS}
}

@article{vsmejkal2022beyond,
  title={Beyond conventional ferromagnetism and antiferromagnetism: A phase with nonrelativistic spin and crystal rotation symmetry},
  author={{\v{S}}mejkal, Libor and Sinova, Jairo and Jungwirth, Tomas},
  journal={Phys. Rev. X},
  volume={12},
  number={3},
  pages={031042},
  year={2022},
  publisher={APS}
}

@article{shao2021spin,
  title={Spin-neutral currents for spintronics},
  author={Shao, Ding-Fu and Zhang, Shu-Hui and Li, Ming and Eom, Chang-Beom and Tsymbal, Evgeny Y},
  journal={Nat. Commun.},
  volume={12},
  number={1},
  pages={7061},
  year={2021},
  publisher={Nature Publishing Group UK London}
}

@article{feng2022anomalous,
  title={An anomalous Hall effect in altermagnetic ruthenium dioxide},
  author={Feng, Zexin and Zhou, Xiaorong and {\v{S}}mejkal, Libor and Wu, Lei and Zhu, Zengwei and Guo, Huixin and Gonz{\'a}lez-Hern{\'a}ndez, Rafael and Wang, Xiaoning and Yan, Han and Qin, Peixin and others},
  journal={Nat. Electron.},
  volume={5},
  number={11},
  pages={735--743},
  year={2022},
  publisher={Nature Publishing Group UK London}
}

@article{bose2022tilted,
  title={Tilted spin current generated by the collinear antiferromagnet ruthenium dioxide},
  author={Bose, Arnab and Schreiber, Nathaniel J and Jain, Rakshit and Shao, Ding-Fu and Nair, Hari P and Sun, Jiaxin and Zhang, Xiyue S and Muller, David A and Tsymbal, Evgeny Y and Schlom, Darrell G and others},
  journal={Nat. Electron.},
  volume={5},
  number={5},
  pages={267--274},
  year={2022},
  publisher={Nature Publishing Group UK London}
}

@article{bai2022observation,
  title={{Observation of spin splitting torque in a collinear antiferromagnet RuO$_2$}},
  author={Bai, Hua and Han, Lei and Feng, XY and Zhou, YJ and Su, RX and Wang, Qian and Liao, LY and Zhu, WX and Chen, XZ and Pan, Feng and others},
  journal={Phys. Rev. Lett.},
  volume={128},
  number={19},
  pages={197202},
  year={2022},
  publisher={APS}
}

@article{weber2024all,
  title={All optical excitation of spin polarization in d-wave altermagnets},
  author={Weber, Marius and Wust, Stephan and Haag, Luca and Akashdeep, Akashdeep and Leckron, Kai and Schmitt, Christin and Ramos, Rafael and Kikkawa, Takashi and Saitoh, Eiji and Kl{\"a}ui, Mathias and others},
  journal={arXiv preprint arXiv:2408.05187},
  doi={10.48550/arXiv.2408.05187},
  urldate      = {2024-8-9},
  note         = {(accessed: 2026-06-10)}
}

@article{zhou2021crystal,
  title={Crystal chirality magneto-optical effects in collinear antiferromagnets},
  author={Zhou, Xiaodong and Feng, Wanxiang and Yang, Xiuxian and Guo, Guang-Yu and Yao, Yugui},
  journal={Phys. Rev. B},
  volume={104},
  number={2},
  pages={024401},
  year={2021},
  publisher={APS}
}

@article{berlijn2017itinerant,
  title={{Itinerant antiferromagnetism in RuO$_2$}},
  author={Berlijn, Tom and Snijders, Paul C and Delaire, O and Zhou, H-D and Maier, Thomas A and Cao, H-B and Chi, S-X and Matsuda, Masaaki and Wang, Yang and Koehler, Michael R and others},
  journal={Phys. Rev. Lett.},
  volume={118},
  number={7},
  pages={077201},
  year={2017},
  publisher={APS}
}

@article{zhu2019anomalous,
  title={{Anomalous antiferromagnetism in metallic RuO$_2$ determined by resonant x-ray scattering}},
  author={Zhu, ZH and Strempfer, J and Rao, RR and Occhialini, CA and Pelliciari, J and Choi, Y and Kawaguchi, T and You, H and Mitchell, JF and Shao-Horn, Y and others},
  journal={Phys. Rev. Lett.},
  volume={122},
  number={1},
  pages={017202},
  year={2019},
  publisher={APS}
}

@article{hiraishi2024nonmagnetic,
  title={{Nonmagnetic ground state in RuO$_2$ revealed by muon spin rotation}},
  author={Hiraishi, M and Okabe, H and Koda, A and Kadono, R and Muroi, T and Hirai, D and Hiroi, Z},
  journal={Phys. Rev. Lett.},
  volume={132},
  number={16},
  pages={166702},
  year={2024},
  publisher={APS}
}

@article{kessler2024absence,
  title={{Absence of magnetic order in RuO$_2$: insights from $\mu$ SR spectroscopy and neutron diffraction}},
  author={Ke{\ss}ler, Philipp and Garcia-Gassull, Laura and Suter, Andreas and Prokscha, Thomas and Salman, Zaher and Khalyavin, Dmitry and Manuel, Pascal and Orlandi, Fabio and Mazin, Igor I and Valent{\'\i}, Roser and others},
  journal={npj Spintronics},
  volume={2},
  number={1},
  pages={50},
  year={2024},
  publisher={Nature Publishing Group UK London}
}

@article{bai2023efficient,
  title={{Efficient spin-to-charge conversion via altermagnetic spin splitting effect in antiferromagnet RuO$_2$}},
  author={Bai, H and Zhang, YC and Zhou, YJ and Chen, P and Wan, CH and Han, L and Zhu, WX and Liang, SX and Su, YC and Han, XF and others},
  journal={Phys. Rev. Lett.},
  volume={130},
  number={21},
  pages={216701},
  year={2023},
  publisher={APS}
}

@article{guo2024direct,
  title={{Direct and inverse spin splitting effects in altermagnetic RuO$_2$}},
  author={Guo, Yaqin and Zhang, Jing and Zhu, Zengtai and Jiang, Yuan-yuan and Jiang, Longxing and Wu, Chuangwen and Dong, Jing and Xu, Xing and He, Wenqing and He, Bin and others},
  journal={Adv. Sci.},
  volume={11},
  number={25},
  pages={2400967},
  year={2024},
  publisher={Wiley Online Library}
}

@article{wang2024inverse,
  title={{Inverse Spin Hall Effect Dominated Spin-Charge Conversion in (101) and (110)-Oriented RuO$_2$ Films}},
  author={Wang, ZQ and Li, ZQ and Sun, L and Zhang, ZY and He, K and Niu, H and Cheng, J and Yang, M and Yang, X and Chen, G and others},
  journal={Phys. Rev. Lett.},
  volume={133},
  number={4},
  pages={046701},
  year={2024},
  publisher={APS}
}

@article{liao2024separation,
  title={{Separation of Inverse altermagnetic spin-splitting effect from inverse spin Hall effect in RuO$_2$}},
  author={Liao, Ching-Te and Wang, Yu-Chun and Tien, Yu-Cheng and Huang, Ssu-Yen and Qu, Danru},
  journal={Phys. Rev. Lett.},
  volume={133},
  number={5},
  pages={056701},
  year={2024},
  publisher={APS}
}

@article{li2025fully,
  title={{Fully Field-Free Spin-Orbit Torque Switching Induced by Spin Splitting Effect in Altermagnetic RuO$_2$}},
  author={Li, Zhuoyi and Zhang, Zhe and Chen, Yuzhe and Hu, Sicong and Ji, Yingjie and Yan, Yu and Du, Jun and Li, Yao and He, Liang and Wang, Xuefeng and others},
  journal={Adv. Mater.},
  volume={37},
  number={12},
  pages={2416712},
  year={2025},
  publisher={Wiley Online Library}
}

@article{wang2025robust,
  title={{Absence of transport altermagnetic spin-splitting effect in RuO$_2$}},
  author={Wang, Yu-Chun and Shen, Zhe-Yu and Lin, Chia-Hsi and Hsu, Wei-Chih and Chen, You-Sheng and Chin, Yi-Ying and Singh, Akhilesh Kr and Lee, Wei-Li and Chen, Chien-Te and Huang, Ssu-Yen and others},
  journal={Nano Letters},
  volume={26},
  number={7},
  pages={2548--2554},
  year={2026},
  publisher={ACS Publications}
}

@article{roy2022unconventional,
  title={Unconventional spin Hall effects in nonmagnetic solids},
  author={Roy, Arunesh and Guimar{\~a}es, Marcos HD and S{\l}awi{\'n}ska, Jagoda},
  journal={Phys. Rev. Mater.},
  volume={6},
  number={4},
  pages={045004},
  year={2022},
  publisher={APS}
}

@article{jeong2024altermagnetic,
  title={Altermagnetic polar metallic phase in ultrathin epitaxially strained RuO2 films},
  author={Jeong, Seung Gyo and Choi, In Hyeok and Nair, Sreejith and Buiarelli, Luca and Pourbahari, Bita and Oh, Jin Young and Lin, Bonnie YX and LeBeau, James M and Bassim, Nabil and Hirai, Daigorou and others},
  journal={Proc. Natl. Acad. Sci.},
  volume={123},
  number={10},
  pages={e2526641123},
  year={2026},
  publisher={National Academy of Sciences}
}

@article{jeong2025metallicity,
  title={{Metallicity and anomalous Hall effect in epitaxially strained, atomically thin RuO$_2$ films}},
  author={Jeong, Seung Gyo and Lee, Seungjun and Lin, Bonnie and Yang, Zhifei and Choi, In Hyeok and Oh, Jin Young and Song, Sehwan and Lee, Seung wook and Nair, Sreejith and Choudhary, Rashmi and others},
  journal={Proc. Natl. Acad. Sci.},
  volume={122},
  number={24},
  pages={e2500831122},
  year={2025},
  publisher={National Academy of Sciences}
}

@article{smolyanyuk2024fragility,
  title={{Fragility of the magnetic order in the prototypical altermagnet RuO$_2$}},
  author={Smolyanyuk, Andriy and Mazin, Igor I and Garcia-Gassull, Laura and Valent{\'\i}, Roser},
  journal={Phys. Rev. B},
  volume={109},
  number={13},
  pages={134424},
  year={2024},
  publisher={APS}
}

@article{wenzel2025fermi,
  title={{Fermi-liquid behavior of nonaltermagnetic RuO$_2$}},
  author={Wenzel, Maxim and Uykur, Ece and R{\"o}{\ss}ler, Sahana and Schmidt, Marcus and Janson, Oleg and Tiwari, Achyut and Dressel, Martin and Tsirlin, Alexander A},
  journal={Phys. Rev. B},
  volume={111},
  number={4},
  pages={L041115},
  year={2025},
  publisher={APS}
}

@article{kiefer2025crystal,
  title={{Crystal structure and absence of magnetic order in single-crystalline RuO$_2$}},
  author={Kiefer, Lara and Wirth, Felix and Bertin, Alexandre and Becker, Petra and Bohat{\`y}, Ladislav and Schmalzl, Karin and Stunault, Anne and Rodr{\'\i}guez-Velamaz{\'a}n, J Alberto and Fabelo, Oscar and Braden, Markus},
  journal={J. Phys. Condens. Matter.},
  volume={37},
  number={13},
  pages={135801},
  year={2025},
  publisher={IOP Publishing}
}

@article{tschirner2023saturation,
  title={{Saturation of the anomalous Hall effect at high magnetic fields in altermagnetic RuO$_2$}},
  author={Tschirner, Teresa and Ke{\ss}ler, Philipp and Gonzalez Betancourt, Ruben Dario and Kotte, Tommy and Kriegner, Dominik and B{\"u}chner, Bernd and Dufouleur, Joseph and Kamp, Martin and Jovic, Vedran and Smejkal, Libor and others},
  journal={APL Mater.},
  volume={11},
  number={10},
  page={101103},
  year={2023},
  publisher={AIP Publishing}
}

@article{fedchenko2024observation,
  title={{Observation of time-reversal symmetry breaking in the band structure of altermagnetic RuO$_2$}},
  author={Fedchenko, Olena and Min{\'a}r, Jan and Akashdeep, Akashdeep and D’Souza, Sunil Wilfred and Vasilyev, Dmitry and Tkach, Olena and Odenbreit, Lukas and Nguyen, Quynh and Kutnyakhov, Dmytro and Wind, Nils and others},
  journal={Sci. adv.},
  volume={10},
  number={5},
  pages={eadj4883},
  year={2024},
  publisher={American Association for the Advancement of Science}
}

@article{liu2024absence,
  title={{Absence of altermagnetic spin splitting character in rutile oxide RuO$_2$}},
  author={Liu, Jiayu and Zhan, Jie and Li, Tongrui and Liu, Jishan and Cheng, Shufan and Shi, Yuming and Deng, Liwei and Zhang, Meng and Li, Chihao and Ding, Jianyang and others},
  journal={Phys. Rev. Lett.},
  volume={133},
  number={17},
  pages={176401},
  year={2024},
  publisher={APS}
}

@article{song2025spin,
  title={{Spin-Orbit Coupling Driven Magnetic Response in Altermagnetic RuO$_2$}},
  author={Song, Jeongkeun and Lee, Seung Hun and Kang, San and Kim, Donghan and Jeong, Ji Hwan and Oh, Taekoo and Lee, Sangjae and Lee, Suyoung and Lee, Sangmin and Ahn, Kyo-Hoon and others},
  journal={Small},
  volume={21},
  number={3},
  pages={2407722},
  year={2025},
  publisher={Wiley Online Library}
}

@article{ahn2019antiferromagnetism,
  title={{Antiferromagnetism in RuO$_2$ as d-wave Pomeranchuk instability}},
  author={Ahn, Kyo-Hoon and Hariki, Atsushi and Lee, Kwan-Woo and Kune{\v{s}}, Jan},
  journal={Phys. Rev. B},
  volume={99},
  number={18},
  pages={184432},
  year={2019},
  publisher={APS}
}

@article{gonzalez2021efficient,
  title={Efficient electrical spin splitter based on nonrelativistic collinear antiferromagnetism},
  author={Gonz{\'a}lez-Hern{\'a}ndez, Rafael and {\v{S}}mejkal, Libor and V{\`y}born{\`y}, Karel and Yahagi, Yuta and Sinova, Jairo and Jungwirth, Tom{\'a}{\v{s}} and {\v{Z}}elezn{\`y}, Jakub},
  journal={Phys. Rev. Lett.},
  volume={126},
  number={12},
  pages={127701},
  year={2021},
  publisher={APS}
}

@article{kubo1,
  title = {Spin-Polarized Current in Noncollinear Antiferromagnets},
  author = {\ifmmode \check{Z}\else \v{Z}\fi{}elezn{\'y}, Jakub and Zhang, Yang and Felser, Claudia and Yan, Binghai},
  journal = {Phys. Rev. Lett.},
  volume = {119},
  issue = {18},
  pages = {187204},
  numpages = {7},
  year = {2017},
  month = {Nov},
  publisher = {American Physical Society},
  doi = {10.1103/PhysRevLett.119.187204},
}

@article{kubo4,
  title = {{Spin-orbit torques in Co/Pt(111) and Mn/W(001) magnetic bilayers from first principles}},
  author = {Freimuth, Frank and Bl\"ugel, Stefan and Mokrousov, Yuriy},
  journal = {Phys. Rev. B},
  volume = {90},
  issue = {17},
  pages = {174423},
  numpages = {10},
  year = {2014},
  month = {Nov},
  publisher = {American Physical Society},
  doi = {10.1103/PhysRevB.90.174423},
}

@article{vzelezny2017spin,
  title={Spin-orbit torques in locally and globally noncentrosymmetric crystals: Antiferromagnets and ferromagnets},
  author={{\v{Z}}elezn{\`y}, J and Gao, H and Manchon, Aur{\'e}lien and Freimuth, Frank and Mokrousov, Yuriy and Zemen, J and Ma{\v{s}}ek, J and Sinova, Jairo and Jungwirth, T},
  journal={Phys. Rev. B},
  volume={95},
  number={1},
  pages={014403},
  year={2017},
  publisher={APS}
}

@article{yang2025large,
  title={{Large Spin-Orbit Torque with Multi-Directional Spin Components in Ni$_4$W}},
  author={Yang, Yifei and Lee, Seungjun and Chen, Yu-Chia and Jia, Qi and Dixit, Brahmdutta and Sousa, Duarte and Odlyzko, Michael and Garcia-Barriocanal, Javier and Yu, Guichuan and Haugstad, Greg and others},
  journal={Adv. Mater.},
  volume={95},
  number={32},
  pages={2416763},
  year={2025},
  publisher={Wiley Online Library}
}

@article{macneill2017control,
  title={{Control of spin--orbit torques through crystal symmetry in WTe$_2$/ferromagnet bilayers}},
  author={MacNeill, D and Stiehl, GM and Guimaraes, MHD and Buhrman, RA and Park, J and Ralph, DC},
  journal={Nat. Phys.},
  volume={13},
  number={3},
  pages={300--305},
  year={2017},
  publisher={Nature Publishing Group UK London}
}

@article{liu2021symmetry,
  title={Symmetry-dependent field-free switching of perpendicular magnetization},
  author={Liu, Liang and Zhou, Chenghang and Shu, Xinyu and Li, Changjian and Zhao, Tieyang and Lin, Weinan and Deng, Jinyu and Xie, Qidong and Chen, Shaohai and Zhou, Jing and others},
  journal={Nature Nanotechnology},
  volume={16},
  number={3},
  pages={277--282},
  year={2021},
  publisher={Nature Publishing Group UK London}
}

@article{jeong2025anisotropic,
  title={{Anisotropic strain relaxation-induced directional ultrafast carrier dynamics in RuO$_2$ films}},
  author={Jeong, Seung Gyo and Choi, In Hyeok and Lee, Seungjun and Oh, Jin Young and Nair, Sreejith and Lee, Jae Hyuck and Kim, Changyoung and Seo, Ambrose and Choi, Woo Seok and Low, Tony and others},
  journal={Sci. Adv.},
  volume={11},
  number={26},
  pages={eadw7125},
  year={2025},
  publisher={American Association for the Advancement of Science}
}

@article{Forte2025,
      title={{Strain Engineering of Altermagnetic Symmetry in Epitaxial RuO$_2$ Films}}, 
  author={Forte, Johnathas DS and Jeong, Seung Gyo and Santhosh, Anand and Lee, Seungjun and Jalan, Bharat and Low, Tony},
  journal={arXiv preprint arXiv:2510.26581},
  doi = {10.48550/arXiv.2510.26581},
  note={(accessed 2026-06-10)}
  }

@article{Yichen2025,
      title={{Observation of mirror-odd and mirror-even spin texture in ultra-thin epitaxially-strained RuO$_2$ films}}, 
      author={Yichen Zhang and Seung Gyo Jeong and Luca Buiarelli and Seungjun Lee and Yucheng Guo and Jiaqin Wen and Hang Li and Sreejith Nair and In Hyeok Choi and Zheng Ren and Ziqin Yue and Alexei Fedorov and Sung-Kwan Mo and Junichiro Kono and Jong Seok Lee and Tony Low and Turan Birol and Rafael M. Fernandes and Milan Radovic and Bharat Jalan and Ming Yi},
      journal={arXiv preprint arXiv:2509.16361},
      doi = {10.48550/arXiv.2509.16361},
      note={(accessed 2025-06-10)}
}

@article{ryden1970electrical,
  title={{Electrical Transport Properties of IrO$_2$ and RuO$_2$}},
  author={Ryden, WD and Lawson, AW and Sartain, Carl C},
  journal={Phys. Rev. B},
  volume={1},
  number={4},
  pages={1494},
  year={1970},
  publisher={APS}
}

@article{guo2008intrinsic,
  title={Intrinsic spin Hall effect in platinum: First-principles calculations},
  author={Guo, Guang-Yu and Murakami, Shuichi and Chen, T-W and Nagaosa, Naoto},
  journal={Phys. Rev. Lett.},
  volume={100},
  number={9},
  pages={096401},
  year={2008},
  publisher={APS}
}

@article{PhysRevB.95.235104,
  title = {Dirac nodal lines and induced spin Hall effect in metallic rutile oxides},
  author = {Sun, Yan and Zhang, Yang and Liu, Chao-Xing and Felser, Claudia and Yan, Binghai},
  journal = {Phys. Rev. B},
  volume = {95},
  issue = {23},
  pages = {235104},
  numpages = {7},
  year = {2017},
  month = {Jun},
  publisher = {American Physical Society},
  doi = {10.1103/PhysRevB.95.235104},
}

@article{noh2025tunneling,
  title={{Tunneling Magnetoresistance in Altermagnetic RuO$_2$-Based Magnetic Tunnel Junctions}},
  author={Noh, Seunghyeon and Kim, Gye-Hyeon and Lee, Jiyeon and Jung, Hyeonjung and Seo, Uihyeon and So, Gimok and Lee, Jaebyeong and Lee, Seunghyun and Park, Miju and Yang, Seungmin and others},
  journal={Phys. Rev. Lett.},
  volume={134},
  number={24},
  pages={246703},
  year={2025},
  publisher={APS}
}

@article{jeong2025magnetic,
title={Magnetic and crystal symmetry control on spin Hall conductivity in altermagnets},
  author={Jeong, Dameul and Kang, Seoung-Hun and Kwon, Young-Kyun},
  journal={Adv. Sci.},
  volume={13},
  number={13},
  pages={e15002},
  year={2026},
  publisher={Wiley Online Library}
}

@Article{Jung2025,
author={Jung, Hyeonjung
and So, Gimok
and Noh, Seunghyeon
and Kim, Gye-Hyeon
and Lee, Jiyeon
and Lee, Jaebyeong
and Lee, Seunghyun
and Seo, Uihyeon
and Han, Dong-Soo
and Oh, Yoon Seok
and Jin, Hosub
and Sohn, Changhee
and Yoo, Jung-Woo},
title={{Reversible Spin Splitting Effect in Altermagnetic RuO$_2$ Thin Films}},
journal={Nano Lett.},
year={2025},
month={Dec},
day={10},
publisher={American Chemical Society},
volume={25},
number={49},
pages={16985-16991},
issn={1530-6984},
doi={10.1021/acs.nanolett.5c03644},
}

@article{Ho2025,
  title = {{Symmetry-breaking induced surface magnetization in nonmagnetic ${\mathrm{RuO}}_{2}$}},
  author = {Ho, Dai Q. and To, D. Quang and Hu, Ruiqi and Bryant, Garnett W. and Janotti, Anderson},
  journal = {Phys. Rev. Mater.},
  volume = {9},
  issue = {9},
  pages = {094406},
  numpages = {11},
  year = {2025},
  month = {Sep},
  publisher = {American Physical Society},
  doi = {10.1103/6fxv-153y},
  }

@article{akashdeep2025surface,
  title={{Surface-Localized Magnetic Order in RuO$_2$ Thin Films Revealed by Low-Energy Muon Probes}},
  author={Akashdeep, Akashdeep and Krishnia, Sachin and Ha, Jae-Hyun and An, Siyeon and Gaerner, Maik and Prokscha, Thomas and Suter, Andreas and Janka, Gianluca and Reiss, G{\"u}nter and Kuschel, Timo and others},
  journal={Appl. Phys. Lett.},
  volume={128},
  number={2},
  page={022406},
  year={2026},
  publisher={AIP Publishing}
}

@article{liu2012spin,
  title={Spin-torque switching with the giant spin Hall effect of tantalum},
  author={Liu, Luqiao and Pai, Chi-Feng and Li, Y and Tseng, HW and Ralph, DC and Buhrman, RA},
  journal={Science},
  volume={336},
  number={6081},
  pages={555--558},
  year={2012},
  publisher={American Association for the Advancement of Science}
}

@article{PhysRevLett.106.036601,
  title = {Spin-Torque Ferromagnetic Resonance Induced by the Spin Hall Effect},
  author = {Liu, Luqiao and Moriyama, Takahiro and Ralph, D. C. and Buhrman, R. A.},
  journal = {Phys. Rev. Lett.},
  volume = {106},
  issue = {3},
  pages = {036601},
  numpages = {4},
  year = {2011},
  month = {Jan},
  publisher = {American Physical Society},
  doi = {10.1103/PhysRevLett.106.036601},
}

@article{quarterman2018demonstration,
  title={{Demonstration of Ru as the 4th ferromagnetic element at room temperature}},
  author={Quarterman, P and Sun, Congli and Garcia-Barriocanal, Javier and Dc, Mahendra and Lv, Yang and Manipatruni, Sasikanth and Nikonov, Dmitri E and Young, Ian A and Voyles, Paul M and Wang, Jian-Ping},
  journal={Nat. commun.},
  volume={9},
  number={1},
  pages={2058},
  year={2018},
  publisher={Nature Publishing Group UK London}
}

@article{ji2013strain,
  title={{Strain induced giant magnetism in epitaxial Fe$_{16}$N$_2$ thin film}},
  author={Ji, Nian and Lauter, Valeria and Zhang, Xiaowei and Ambaye, Hailemariam and Wang, Jian-Ping},
  journal={Appl. Phys. Lett.},
  volume={102},
  number={7},
  year={2013},
  publisher={AIP Publishing}
}

\clearpage\newpage

\end{document}